\documentclass[prd]{revtex4}
\begin{document}

\title{The need for dark matter in galaxies}
\author{David Garfinkle}
\email{garfinkl@oakland.edu}
\affiliation{Department of Physics, Oakland University, Rochester, MI 48309}

\begin{abstract}

Cooperstock and Tieu have proposed a model to account for 
galactic rotation curves 
without invoking dark matter.  I argue that no model of 
this type can work.  

\end{abstract}
\maketitle

In a recent paper\cite{CT} Cooperstock and Tieu have proposed a model of 
galaxies that involves only ordinary matter.  The model of Ref. \cite{CT} is
supposed to account for galactic rotation curves through the non-linearity
of the Einstein field equations.  The specific model of Ref. \cite{CT} 
was shown
in Refs. \cite{MK,VL} to be incorrect in the sense 
that it contains extra matter. 
This matter is essentially a domain wall in the plane of the galaxy.   
Nonetheless, one might wonder whether the model of Ref. \cite{CT} could be 
somehow fixed, that is whether the non-linearity of Einstein's equation
could be a viable mechanism for explaining galactic rotation curves
without dark matter.  I argue that no such mechanism is possible: the 
argument has two pieces: one having to do with the post-Newtonian 
approximation and the other having to do with the non-existence of geons.

The post-Newtonian approximation is usually presented as an expansion 
of the Einstein field equations in powers of $1/c$.  However it can also
be viewed as an iterative procedure that begins with a solution of the
equations of Newtonian gravity and produces a sequence of better and better
approximations to a solution to the Einstein field equations.\cite{dire}  The 
post-Newtonian approximation defines from the metric $g_{\mu \nu}$ 
the quantity ${h^{\mu \nu}} = {\eta ^{\mu \nu}} - {{(-g)}^{1/2}}
{g^{\mu \nu}}$ where $g$ is the determinant of $g_{\mu \nu}$ and
$\eta _{\mu \nu}$ is the metric of flat spacetime.  The full 
non-linear Einstein
field equations are then written in the form
\begin{equation}
{\partial _\alpha}{\partial ^\alpha} {h^{\mu \nu}} = 
- 16 \pi {\tau ^{\mu \nu}} 
\label{EFE}
\end{equation}  
Here ${\partial _\alpha}{\partial ^\alpha} $ is the flat spacetime 
wave operator and the quantity 
$\tau ^{\mu \nu}$ is essentially 
an effective stress-energy tensor containing a contribution from the ordinary
stress-energy tensor of matter and another contribution from the
nonlinear terms in the Einstein field equations.  Since the
flat spacetime wave operator can be inverted using a Green's function, 
given a guess about the metric and matter fields, equation (\ref{EFE}) can 
be used to find an improved guess for the metric.  The matter equations
of motion can similarly be iterated to provide an improved guess for the
matter fields.  This, along with an approximation for
slow motion forms the basis of the post-Newtonian approximation.  For our
purposes what is important is that the post-Newtonian approximation does
take into account the non-linearity of the Einstein field equations, and
that for situations that are slow motion and weak gravity, the corrections
to the Newtonian solution are small.  Since galaxies certainly appear to 
be slow motion weak gravity systems (except for the black hole in the 
center) one would expect that the answer found by the post-Newtonian 
approximation should both appropriately take into account the non-linearity
of the Einstein field equations and approximately agree with the results
of Newtonian theory.  In other words, since a Newtonian treatment of 
a galaxy reveals the presence of dark matter, so does a general relativistic
treatment.  

The only way around this argument would be if somehow the galaxy were not
weakly gravitating.  Then the ``dark matter'' would be provided by the
effective energy density of the gravitational field 
and that field would have to be treated
in full general relativity rather than the post Newtonian approximation.
The matter of a galaxy could not hold together
such a strong gravity configuration, so for this picture to work, the
gravity waves would have to hold themselves together.  In other words, the
galaxy would effectively be a geon, a vacuum configuration that 
holds itself together:
the dark matter would simply be the effective energy density of the
gravitational field and the ordinary matter would simply be an additional
component of the galaxy held in mostly by the gravitational field of the 
geon.  However, this does not work either.  Exactly stationary geons do
not exist because the Komar formula\cite{komar} 
for the mass of a stationary system 
yields zero for vacuum solutions.  But configurations that are not 
stationary tend to emit gravitational radiation and settle down to 
stationary states.  Furthermore the theorem of Christodoulou and 
Klainerman\cite{dc} states that vacuum systems that are not 
sufficiently strongly gravitating will disperse.  So the geon would have to be
sufficiently strong to evade the theorem of Ref. \cite{dc} and at the same
time sufficiently weak so that the stars in the galaxy are on slow 
moving orbits.  Given this plethora of needed but opposing conditions,
we can be confident that no such geon exists.

In conclusion, the non-linearity of Einstein's equations will not save us
from the necessity of dark matter.

\section{Acknowledgements}

I would like to thank Bob Wald for helpful discussions.  
This work was partially supported by
NSF grant PHY-0456655 to Oakland University.

\end{document}